\RequirePackage{amsmath}
\documentclass[runningheads]{llncs}

\usepackage[T1]{fontenc}

\usepackage{cite}
\usepackage{amssymb,amsfonts}
\usepackage{graphicx}
\usepackage{textcomp}
\usepackage{xcolor}
\def\BibTeX{{\rm B\kern-.05em{\sc i\kern-.025em b}\kern-.08em
    T\kern-.1667em\lower.7ex\hbox{E}\kern-.125emX}}
\usepackage{hyperref}
\usepackage{todonotes}

\usepackage{enumitem}

\newcommand{\Astree}{Astr{\'e}e}
\newcommand{\ISO}{ISO~26262}
\newcommand{\CPP}{C\nolinebreak\texttt{++}}	
\newcommand{\FC}{\mathit{FC}}
\newcommand{\VC}{\mathit{VC}}

\usepackage{tikz}
\usetikzlibrary{calc,positioning,shapes}
\usetikzlibrary{arrows}
\usetikzlibrary{backgrounds}

\usepackage{listings}

\usepackage[caption=false]{subfig}
\usepackage{colortbl}

\usepackage{lmodern}	

\usepackage{arydshln}	
\usepackage{changepage}

\usepackage[ruled,linesnumbered]{algorithm2e}

\SetCommentSty{mycommfont}

\SetNlSty{textbf}{\color{black}}{}

\definecolor{myBlue}{rgb}{0.0,0.355,0.652}
\definecolor{myBlueBackground}{rgb}{0.6, 0.81, 0.93}
\definecolor{myGreen}{rgb}{0.390,0.695,0.285}
\definecolor{myRed}{rgb}{0.7,0,0}
\definecolor{darkcyan}{rgb}{0.0, 0.55, 0.55}

\begin{document}

\title{Formal Runtime Error Detection During Development in the Automotive Industry}

\titlerunning{Formal Runtime Error Detection in the Automotive Industry}
%

\author{Jesko Hecking-Harbusch
\and
Jochen Quante
\and
Maximilian Schlund
}
\authorrunning{J.\ Hecking-Harbusch, J.\ Quante, and M.\ Schlund}
%
\institute{Bosch Research, Renningen, Germany\\
\email{firstname.lastname@bosch.com}}

\maketitle              
%


\begin{abstract}
	Modern automotive software is highly complex and consists of millions lines of code.
    For safety-relevant automotive software, it is recommended to use sound static program analysis to prove the absence of runtime errors. 
    However, the analysis is often perceived as burdensome by developers because it runs for a long time and produces many false alarms.
	If the analysis is performed on the integrated software system, there is a scalability problem, and the analysis is only possible at a late stage of development.
	If the analysis is performed on individual modules instead, this is possible at an early stage of development, but the usage context of modules is missing, which leads to too many false alarms.

    In this case study, we present how automatically inferred contracts add context to module-level analysis.
	Leveraging these contracts with an off-the-shelf tool for abstract interpretation makes module-level analysis more precise and more scalable.
    We evaluate this framework quantitatively on industrial case studies from different automotive domains.
    Additionally, we report on our qualitative experience for the verification of large-scale embedded software projects.

\end{abstract}


\section{Introduction}
\label{sec:intro}

A vehicle comprises many software components that are relevant for its safety.
Think about the electronic stability control, which improves the stability of a vehicle by automatically applying the brakes at a wheel to avoid skidding.
A malfunction of such a system blocking a single wheel could result in danger for the occupants and other traffic participants. 
To prevent such incidents, the quality of the deployed embedded software is of great importance.
The standard \ISO{}~\cite{iso26262} comprises the state of the art of technologies to ensure the \emph{functional safety} of electronic systems including their software in road vehicles.
Functional safety requires the ``absence of unreasonable risk due to hazards caused by malfunctioning behavior of electrical/electronic systems''~\cite{iso26262}.

A possible reason for software malfunctions is the occurrence of runtime errors, e.g., null pointer dereferences or array-out-of-bounds accesses.
In languages like C and \CPP{}, runtime errors lead to undefined behavior, which is disastrous in terms of functional safety.
\ISO{} recommends \emph{abstract interpretation}~\cite{cousot:popl77,rival:book20} to prove the absence of runtime errors because it is sound and can be scalable.
Abstract interpretation provides a general framework based on the theory of semantic abstraction to analyze software.
Concrete states of the software are overapproximated by finitely many abstract states such that a fixpoint can be calculated in a finite number of steps.
This fixpoint overapproximates all behavior of the software and is checked for runtime errors.

\emph{Soundness} is an essential property of abstract interpretation, which requires that all runtime errors are found. 
So, if none are found, then the absence of runtime errors is proven.
However, soundness comes at the price of possibly getting more runtime errors reported than there actually are.
This means that in addition to the \emph{true alarms} reported by abstract interpretation, which represent runtime errors, there can be \emph{false alarms} that do not represent runtime errors but are caused by the overapproximation in abstract interpretation.

Commercial tools like \Astree{}~\cite{astree}, Polyspace Code Prover~\cite{polyspaceCodeProver}, and TrustInSoft Analyzer~\cite{trustInSoftAnalyzer} enable the usage of abstract interpretation in industrial contexts~\cite{DBLP:conf/safecomp/SouyrisD07,DBLP:journals/fmsd/CousotCFMMR09,duprat2016spreading,kaestner2017finding,DBLP:conf/icse/TodorovBT18}.
They are used for verifying safety-critical software in the aerospace industry, the automotive industry, and other industrial domains.

There are two ways to apply abstract interpretation in an industrial context: It can be applied to each isolated module separately, or on the integrated system consisting of all implemented modules and their interactions.
The first approach can be performed in early development stages, but the usage context of modules is missing because other modules are not implemented yet.
This leads to many false alarms.
For the second approach, the analysis is only possible at a late stage of development when fixing errors causes enormous costs~\cite{boehm:81}.
Furthermore, its runtime might not scale well with the size of the software. For such systems, the analysis can only be performed with low precision settings of the analyzer leading again to many false alarms.
In general, static program analysis is perceived as burdensome when taking too long, producing too many false alarms, or causing considerable additional manual effort~\cite{DBLP:conf/icse/JohnsonSMB13,DBLP:conf/kbse/ChristakisB16,DBLP:journals/tse/DoWA22}.

In this case study, we show how to make abstract interpretation applicable to individual modules of automotive software early during development without the downsides mentioned above.
Applying abstract interpretation early during development is the most efficient way to handle alarms as developers get feedback quickly and still know the code when evaluating and possibly fixing alarms~\cite{yamaguchi:vmcai19}.

We develop a general framework around an underlying tool for abstract interpretation.
Our framework prepares the abstract interpretation task such that the tool for abstract interpretation can analyze it with higher precision.
To this end, our framework automatically generates a \emph{verification harness} for a module under analysis.
The verification harness contains the module under analysis and additional code to check all functions and their interplay with meaningful inputs. 
We use \Astree{}~\cite{blanchet:book02,blanchet:pldi03,kaestner:erts10} as the underlying tool for abstract interpretation.
The core concepts of our framework can be easily transferred to other tools for abstract interpretation, but their technical realization might differ. 

When generating the verification harness early during development, we rely on \emph{contracts} at the interfaces of modules.
Contracts enrich the syntactic interface between modules by semantic annotations.
They can define that pointers represent arrays of certain lengths and can constrain the domain of variables. 
They can also state preconditions and postconditions of functions and whether these functions are called for initialization or cyclically in an embedded system.

A further core component of our framework is the \emph{automatic inference} of contracts.
This is the only viable option to obtain contracts for large legacy code bases that cannot be manually annotated as this would cause excessive effort.
For automatic inference of contracts, we use different sources of information.
First, we translate information from interface descriptions of modules into contracts.
Second, we utilize information from abstract interpretation results of previously analyzed modules.
We propagate information over all modules and iterate until a fixpoint is reached.
During this iterative refinement, the contracts become more and more precise, and thus the number of found alarms can decrease significantly. 

We evaluate our framework on real-world projects from the automotive domains of driver assistance, braking, and cruise control and show that it reduces the number of found alarms significantly.
For example, for the domain of braking, our framework decreases the number of alarms by approx.\ 50\% while increasing code coverage by approx.\ 50\%.
By using contracts in module-level abstract interpretation, our framework makes it also possible for developers to use abstract interpretation early during development and benefit from low analysis runtimes for individual modules. 
Due to the automatic inference of contracts, they rarely need to add contracts manually.
Thus, our framework enables developers to analyze modules in projects that were not analyzable in reasonable time before.

When deploying our framework in different projects from different business units, we learned that automation is key in successfully using abstract interpretation.
Developers often lack time and need to justify how to spend it. 
Here, automation in the form of our framework steps in.
People can get first results after a few days with our framework in comparison to needing up to a month to being able to use \Astree{} properly.
We designed our framework to be adaptable to new projects with custom extensions to account for the specific architecture models, processes, and conventions used in the projects.
The first results are often already helpful for developers and make them want to continue to use our framework.
Developers appreciate the short runtimes of module-level analysis because they do not block their usual workflows.
With our framework, developers can quickly benefit from years of experience in how to deploy abstract interpretation in the automotive domain.

Overall, this paper shows how to use abstract interpretation to detect runtime errors in real-world automotive software.
Our key contributions are
\begin{itemize}
  \item bringing together abstract interpretation and contracts to enable the verification of automotive software early during development,
  \item techniques for automatically inferring contracts of automotive software to minimize manual effort for developers,
  \item a quantitative evaluation of our framework of module-level abstract interpretation with automatically inferred contracts on three case studies from the automotive domain, and
  \item a qualitative report on our experiences of using our framework on large-scale production code early during development.
\end{itemize}


\section{Background}
\label{sec:background}

\begin{figure}[t]%
    \centering
    \begin{minipage}{0.825\linewidth}
    \subfloat[\label{fig:AIprocess}]{%
    \resizebox{1\textwidth}{!}{\begin{tikzpicture}[		
    ->,
    very thick,
    >=stealth',
    node distance=15mm and 45mm,
    lbl/.style={
       align=center
    },
    elem/.style={
        rectangle,
        rounded corners,
        draw=black, 
        very thick,
        text centered,
        minimum height=10mm,
        minimum width=22mm,
    }
]
\tikzstyle{myarrows}=[line width=1mm,-triangle 45,postaction={line width=2mm, shorten >=2mm, -}]

\node[elem, bottom color=myGreen!50, top color=white, align=center] (concreteBefore) {Concrete\\Domain\\\emph{int}};
\node[elem, below=of concreteBefore,bottom color=myBlue!50, top color=white, align=center] (abstractBefore) {Abstract\\Domain\\\emph{zero domain}};

\node[elem, right=of concreteBefore, bottom color=myGreen!50, top color=white, align=center] (concreteAfter) {Concrete\\Domain\\\emph{int}};
\node[elem, below=of concreteAfter, bottom color=myBlue!50, top color=white, align=center] (abstractAfter) {Abstract\\Domain\\\emph{zero domain}};

\draw[very thick] (concreteBefore) -- node[above] {Operational Semantics} node[below] {\emph{int addition}} (concreteAfter);
\draw[very thick] (abstractBefore) -- node[above] {Abstract Semantics} node[below] {\emph{zero domain addition}} (abstractAfter);
\draw[very thick, dotted] (concreteBefore) -- node[left] {Abstraction} node[right] {\footnotesize{$\text{\emph{value}} \rightarrow \begin{cases}
    {=} 0     & \text{if \emph{value}} = 0 \\
    {\neq} 0  & \text{otherwise}
\end{cases}$}} (abstractBefore);
\draw[very thick, dotted] (abstractAfter) -- node[left] {Concretization} node[right] {\footnotesize{$\text{\emph{value}} \rightarrow \begin{cases}
    0                       & \text{if \emph{value}} = {=} 0 \\
    \text{any int} \neq 0   & \text{if \emph{value}} = {\neq} 0 \\
    \text{any int}          & \text{otherwise}
\end{cases}$}} (concreteAfter);
\end{tikzpicture}}%
    }
    \end{minipage}%
	\begin{minipage}{0.025\linewidth}
		~
	\end{minipage}%
    \begin{minipage}{0.15\linewidth}
		\centering
	    \subfloat[\label{fig:lattice}]{%
   		\resizebox{0.85\textwidth}{!}{\begin{tikzpicture}[		
    ->,
    >=stealth',
    node distance=7.5mm and 7.5mm,
    lbl/.style={
       align=center
    },
    elem/.style={
        text centered,
    }
]
\tikzstyle{myarrows}=[line width=1mm,-triangle 45,postaction={line width=2mm, shorten >=2mm, -}]

\node[] (bot) {$\bot$};
\node[above of=bot, left of=bot] (left) {${=}0$};
\node[above of=bot, right of=bot] (right) {${\neq}0$};
\node[above of=bot, above of=bot] (top) {$\top$};

\draw[] (bot) -- (left);
\draw[] (bot) -- (right);
\draw[] (left) -- (top);
\draw[] (right) -- (top);
\end{tikzpicture}}%
	   	}\\
	    \subfloat[\label{fig:lattice:addition}]{%
	    \resizebox{0.85\textwidth}{!}{\begin{tabular}{>{\columncolor[gray]{0.8}}c | c | c | c | c}
    \rowcolor[gray]{0.8}$+$ & $\bot$ & ${=}0$ & ${\neq}0$ & \,$\top$\, \\
    \hline
    $\bot$ & $\bot$ & ${=}0$ & ${\neq}0$ & $\top$ \\
    \hline
    ${=}0$ & ${=}0$ & ${=}0$ & ${\neq}0$ & $\top$ \\
    \hline
    ${\neq}0$ & ${\neq}0$ & ${\neq}0$ & $\top$ & $\top$ \\
    \hline
    $\top$ & $\top$ & $\top$ & $\top$ & $\top$
\end{tabular}
}%
    }
    \end{minipage}
    \caption{Example of abstract interpretation for addition of two integers: The general idea is illustrated in \ref{fig:AIprocess}. The \emph{zero}-lattice shown in \ref{fig:lattice} tracks whether a value is zero or not. The abstract semantics of the addition operator in the zero domain is given in \ref{fig:lattice:addition}.}%
    \label{fig:abstract-interpretation}%
\end{figure}
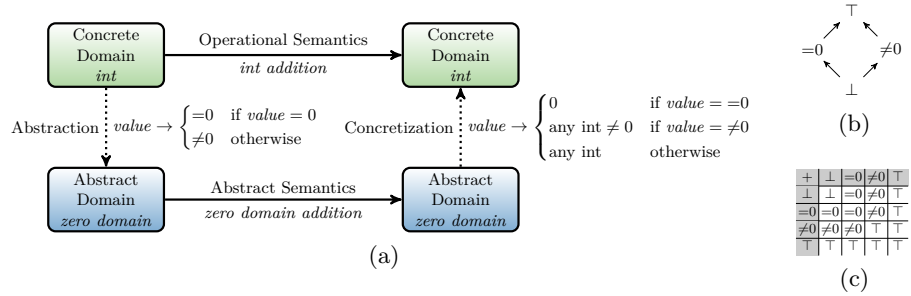

\emph{Abstract interpretation}~\cite{cousot:popl77,rival:book20} is a technique for static program analysis, i.e., the code is analyzed without executing it. 
The idea is to interpret the given code with \emph{abstract values}. 
Abstract values can come for example from the \emph{abstract domains} of intervals, linear equalities, or congruence relations on integers. 
Abstract values are elements of complete lattices.
A \emph{complete lattice} is a partially ordered set in which all subsets have both a \emph{greatest lower bound} $\sqcap$ and a \emph{least upper bound} $\sqcup$. The latter are generalizations of intersection and union on sets. The maximum and minimum elements~$\top$ and~$\bot$ of the entire lattice are called top and bottom, respectively.
\emph{Abstract operations} on abstract values are monotonic functions. 
By repeatedly calculating the abstract operations for all instructions of the code, a fixpoint is reached after some iterations due to the monotonicity property. 
This ensures sound overapproximation of all possible concrete values.

\begin{example} 
	In Fig.~\ref{fig:abstract-interpretation}, we perform an addition of two integer variables that both are abstracted by the zero domain, which tracks whether an integer variable is zero (${=}0$) or nonzero (${\neq}0$).
	Fig.~\ref{fig:AIprocess} shows the abstraction and concretization functions.
	The lattice for the zero domain is given in Fig.~\ref{fig:lattice} where the maximum element~$\top$ encodes that the variable can be zero and nonzero, and the minimum element~$\bot$ encodes that the variable is neither zero nor nonzero.
	The minimum element can be helpful to encode that a certain variable is not reachable or not initialized.
	Abstraction of a concrete integer value is straightforward by using the appropriate member of the zero domain.
	Addition in the zero domain is defined by the table in Fig.~\ref{fig:lattice:addition}.
	For example, the sum of two variables equal to zero is also zero, whereas the sum of two variables different from zero can either be zero or nonzero and hence is $\top$.
	Notice that calculating the sum of two variables both abstracted by nonzero in the abstract semantics represents the addition of almost all concrete values for the two variables.
\end{example}

Using an adequate abstract domain, abstract interpretation can prove certain properties of software. 
For example, an interval domain may be used to prove that arithmetic overflows cannot happen for given range limitations of the input values. 
Current abstract interpretation tools like \Astree{}~\cite{blanchet:book02,blanchet:pldi03,kaestner:erts10} can scale up to millions of lines of code. 
They guarantee soundness at the price of potentially many false alarms but offer the possibility to add annotations in order to feed the analysis with additional information that can greatly improve precision. 
However, even with a thorough understanding of abstract interpretation, intensive manual work by developers is required to come up with annotations that reduce the number of false alarms. 
To enable module-level analysis, further manual work is required to create \textit{stubs} for emulating missing dependent functions and \textit{drivers} for supplying subject functions with adequate input.

Next, we introduce annotations specific to \Astree{} that will be used throughout the paper.
If they are encountered in code by \Astree{}, these annotations alter abstract values.
The directive \texttt{\_\_ASTREE\_modify((x;full\_range))} changes the abstract value of the given variable~\texttt{x} to~$\top$, i.e., previous restrictions on the abstract value do not hold anymore.
Assuming~\texttt{x} is of type \texttt{unsigned char} and the interval abstract domain is used, \Astree{} continues calculations after the directive with the interval $[0, 255]$ for~\texttt{x}. 
The directive \texttt{\_\_ASTREE\_assert((x > 0))} makes \Astree{} check the given condition on the abstract value.
Executing the previous two annotations in the given order without other instructions in between will result in the assertion failing as the abstract value will contain zero due to being set to~$\top$ before.
The directive \texttt{\_\_ASTREE\_known\_fact((x > 0))} allows defining conditions that are assumed as true to alter the abstract value.
This makes it possible to make abstract values more precise.
For the previous example of setting \texttt{x} to $\top$, this directive can exclude zero from the interval afterwards.

\emph{Contracts} as introduced by Meyer refine interface specifications for software modules~\cite{meyer:book88}. 
For a given function, they define that certain {\em preconditions} must be fulfilled when the function is called and that the function then guarantees certain {\em postconditions} when it returns. 
Below is an example for a \emph{function contract}:

\begin{lstlisting}
/// [[ requires: x >= 0.0 ]]
/// [[ ensures: return >= 0.0 ]]
float sqrt(float x);
\end{lstlisting}
Traditionally, such contracts are checked dynamically by adding assertions at the beginning and end of a function.
In Sec.~\ref{subsec:contracts}, we will see how contracts can be translated into \Astree{} annotations.
Another class of contracts are {\em invariants}. They are two types of invariants: A class invariant specifies that a condition must always hold at interface borders, i.e., it is implicitly added to all preconditions and postconditions of the class. A global invariant is globally valid and thus must be checked whenever a relevant variable is modified.
In this paper, global invariants on single variables are called \emph{variable contracts}.
Contract specifications for \CPP{} are being discussed for inclusion in an upcoming \CPP{}2x version~\cite{std:contracts}.


\section{Module-level Abstract Interpretation with Contracts}
\label{sec:approach}

\begin{figure}[t]
    \centering
    \resizebox{.8\textwidth}{!}{\input{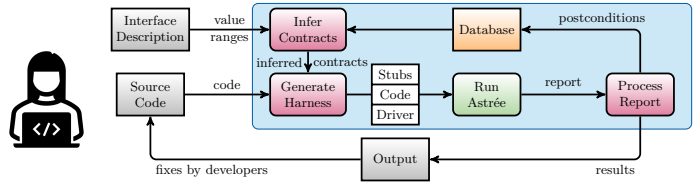}}
    \caption{
        The inner workings of our framework (highlighted in blue) and how developers interact with it are depicted.
        Source code and interface description for software modules are given to our framework, which internally uses \Astree{} for abstract interpretation.
        The results are aggregated over subsequent runs in a database to infer more contracts and output to the developers who then can fix potential runtime errors in their code. 
    }
    \label{fig:overview}
\end{figure}

In \emph{module-level analysis}, we apply abstract interpretation to a given module.
There is some information missing in comparison with \emph{integration analysis} where the entire software project is analyzed in late stages of development.
Module-level analysis requires a starting point (i.e., a main function that invokes the module's functions), initialization of variables with abstract values, and meaningful stubs for missing functions. 
In this section, we describe how to automatically generate the \emph{verification harness}~\cite{DBLP:conf/spin/HolzmannJ04} for module-level analysis based on results of code analysis and how contracts~\cite{meyer:book88} can be used to enhance the harness.

Figure~\ref{fig:overview} shows how the first step called ``Generate Harness'' fits into our framework. 
The code to be checked along with the generated stubs and verification driver are given to \Astree{} for analysis.
The report from the \Astree{} run is processed and interpreted with the knowledge of how the harness was generated. 
This is done to report potential runtime errors and contract violations to the developer and to infer additional contracts for subsequent runs. 
For inferring contracts, relevant abstract values are stored in a database shared across the complete software project.

\begin{algorithm}[t]
\SetKwInOut{Input}{Input}\SetKwInOut{Output}{Output}
\Input{Module $m${\color{orange}, function contracts $\FC$, variable contracts $\VC$}}
\Output{Potential runtime errors in $m$, {\color{darkcyan} return and variable values}}
\BlankLine
perform dependency analysis of $m$\;
\label{line:depAnalysis}
include the code of $m$\;
\label{line:startStubs}
\ForEach{undefined but used function $f$}{
	generate a stub for $f$ {\color{orange}that considers $\FC(f)$}\;
}
\label{line:endStubs}
generate a main function header\;
\tcc{start main function body}
\label{line:startDriver}
\ForEach{used global variable $v$}{
	generate initialization {\color{orange}according to $\VC(v)$}\;
}
create required \CPP{} instances\;
generate an endless loop with random switch\;
\ForEach{function $f$ to be called from driver}{
	generate a distinct case\tcp*[l]{containing:}
	initialization of each argument {\color{orange}according to preconditions from $\FC(f)$}\;
	call of $f$ with these arguments\;
	{\color{orange}postcondition check for $\FC(f)$}\;
	{\color{darkcyan}directives to output return and variable values}\;
	\label{line:additionalDirectives}
}
\tcc{end main function body}
\label{line:endDriver}
{\color{darkcyan}(}{\it alarms}{\color{darkcyan}, {\it values})} = run \Astree{} analysis on this harness\;
\Return \textcolor{darkcyan}{(}{\it alarms}\textcolor{darkcyan}{, {\it values})}\;
\label{line:callAstree}
\BlankLine
\caption{Verification of a Module. \newline {\color{orange}Orange} parts are added when using contracts. \newline {\color{darkcyan}Cyan} parts are added for inferring contracts.}
\label{alg:stubgen}
\end{algorithm}

\subsection{Generating a Verification Harness}
\label{subsec:harnessgen}

To generate a verification harness~\cite{DBLP:conf/spin/HolzmannJ04} for a given module, we perform the following three steps, which are shown as pseudocode in Algorithm~\ref{alg:stubgen}:

\begin{enumerate}[label=(\roman*)]
	\item Dependency analysis (line~\ref{line:depAnalysis}): We identify the complete interface of the module under analysis, i.e., provided and required types, functions, and global variables, which is used in all subsequent steps.

	\item \label{enum:stubs} Stub generation (lines \ref{line:startStubs} to \ref{line:endStubs}): For each function the module under analysis depends upon, we generate a \emph{stub} that returns $\top$ for all return values. Note that \Astree{} automatically generates such a stub for missing functions, but we will build upon this when using contracts in Sec.~\ref{subsec:contracts}.

	\item \label{enum:driver} Verification driver generation (lines \ref{line:startDriver} to \ref{line:endDriver}): We generate an entry point from where the analysis should commence following the basic structure of an embedded system:
	\begin{lstlisting}
initialization();
while(1) { cyclic_tasks(); }
	\end{lstlisting}
	During \texttt{initialization()}, global variables are set to $\top$ and required \CPP{} instances are created. 
	For \texttt{cyclic\_tasks()}, all public functions of the module are called in all possible orders and supplied with all possible inputs.
\end{enumerate}

\begin{example}
	In~\ref{enum:stubs}, the function \texttt{sqrt()} from Sec.~\ref{sec:background} yields the following stub:
	\begin{lstlisting}
float sqrt(float x) {
	float res;
	__ASTREE_modify((res;full_range));
	return res;
}
	\end{lstlisting}
The stub accepts all possible float values and returns all possible float values.
\end{example}

\begin{example}
	Consider a module with a specific initialization function \texttt{void g()}, two public functions \texttt{void f1()} and \texttt{int f2(int x)}, and a global integer variable \texttt{glob\_v}.
	In~\ref{enum:driver}, our framework then generates the following code with comments added for explanatory purposes:
\\ 
\begin{lstlisting}
int glob_v;
// initialize global variables
__ASTREE_modify((glob_v;full_range));
// call the module's init-function
g();
unsigned char decision;
while(1) {
  // non-deterministic choice
  __ASTREE_modify((decision;full_range));
  // decision -> [0, 255]
  switch(decision) {
    case 0: {
      f1();
      break;
    }  
    case 1: {
      int p1;
      // set f2's input to full range
      __ASTREE_modify((p1;full_range));
      int res = f2(p1);
      break;
    }  
  } 
}
\end{lstlisting}
	This analysis driver calls \texttt{f1} and \texttt{f2} after \texttt{g} in all possible call orderings such as \texttt{g, f1, f1, \dots} and \texttt{g, f2, f1, f2, \dots} as \texttt{decision} is set to full range.
\end{example}

\subsection{Contracts and their Effects}
\label{subsec:contracts}

To increase the precision of \Astree{} when analyzing the verification harness, we have to substantiate stubs and driver. 
We use {\em contracts}~\cite{meyer:book88} to do so.
The contracts in our framework are geared towards proving the absence of \emph{runtime errors} using abstract interpretation.
They are not aimed at proving functional properties and are usually not expressive enough to do so. 
However, this lack in expressiveness makes them checkable with low overhead using tools like \Astree{}.
Contracts can be provided to our framework by writing them as comments into header files. 
We illustrate the syntax in the following examples.

\paragraph{Pre- and postconditions:}
Via \texttt{requires}, we can express conditions the inputs of a function need to fulfill, and via \texttt{ensures}, we can express guarantees on its outputs.
This is used for two purposes: If the function is stubbed, the contract refines the stub; if the function is under check, the contract refines the driver.

Remember the two contracts for function \texttt{sqrt()} from Sec.~\ref{sec:background}. Using these contracts, the stub is refined to:

\begin{lstlisting}
float sqrt(float x) {
  float res;
  // precondition:
  __ASTREE_assert((x >= 0.0));
  // postcondition:
  __ASTREE_modify((res;full_range));
  // res -> [-3.4028e38, 3.4028e38]
  __ASTREE_known_fact((res >= 0.0));
  // res -> [0.0, 3.4028e38]
  return res;
}
\end{lstlisting}

This refined stub leads to more precise results for modules that call function \texttt{sqrt()} without knowing its implementation.
The reason for this improvement is that the sign of the return value is now known to the callers.

When checking the function, our verification driver calls it with the input [0.0, FLOAT\_MAX], and the \emph{ensures} condition is translated into assertions that check the implementation for compliance with the contract.
The case calling \texttt{sqrt()} in the verification harness would be:

\begin{lstlisting}
    case 0: {
      float p;
      __ASTREE_modify((p;full_range));
      __ASTREE_known_fact((p >= 0.0));
      float res = sqrt(p);
      __ASTREE_assert((res >= 0.0));
      break;
    }  
\end{lstlisting}

Through this dual use of the contract, our framework a) ensures that an implementation fulfills the contract and b) allows analyzing function invocations with higher precision.
The extensions for considering contracts in Algorithm~\ref{alg:stubgen} are highlighted in orange.
Note that contracts for postconditions can also specify side effects on other variables than the return value, e.g., global variables or arguments passed by reference.

\paragraph{Array specifications:}
In C/\CPP{}, pointers given to functions are often arrays.
Our framework offers the possibility to specify that such a pointer is an array of a certain length.
The right-hand side of this specification can refer to another argument, a global variable or constant, or a simple expression like var $+$ const.

Consider the function \texttt{memcmp} from the C~standard library, which compares the first \texttt{n} bytes of the block of memory pointed to by \texttt{ptr1} to the first \texttt{n} bytes pointed to by \texttt{ptr2}.
A possible contract for this function is:
\begin{lstlisting}
/// [[ arrayspec: length(ptr1) >= n ]]
/// [[ arrayspec: length(ptr2) >= n ]]
int memcmp(const void *ptr1, const void *ptr2, size_t n);
\end{lstlisting}	
In the generated stub, the length check will yield an invalid-dereference alarm in \Astree{} if it does not succeed:
\begin{lstlisting}
int memcmp(const void *ptr1, const void *ptr2, size_t n) {
	((char *)ptr1)[n-1]; // check length
	((char *)ptr2)[n-1]; // check length
	int ret;
	__ASTREE_modify((ret; full_range));
	return ret;
}
\end{lstlisting}
The \texttt{arrayspec} contracts are also considered when creating input data for the corresponding function in the driver.

\paragraph{Invariants:}
Our invariants are expressions on global variables that should always hold.
The example below shows that the counter of the current cylinder is always at most the maximum number of cylinders defined in the project.
\begin{lstlisting}
/// [[ invariant: System_ctCyl <= NUM_CYL ]]
uint8 System_ctCyl;
\end{lstlisting}

We can also specify constraints on members of structs:
\begin{lstlisting}
typedef struct {
	/// [[ invariant: Id <= LED_NUMBER ]]
	uint8 Id;
	[...]
} LED;
\end{lstlisting}
Our framework translates all such invariants into \texttt{\_\_ASTREE\_global\_assert} statements, which makes \Astree{} check the condition each time the respective variable is modified.

\paragraph{Sequence specifications:}
A developer can specify when functions from the public interface of the module under analysis are called.
This can either happen initially once or cyclically afterwards (as introduced in \ref{enum:driver} in Sec.~\ref{subsec:harnessgen}):
\begin{lstlisting}
/// [[ sequence: init ]]
void initialization();
/// [[ sequence: cyclic ]]
void run();
\end{lstlisting}
In the generated driver, init functions are called before the while loop and cyclic functions in the cases within it.
Cyclic functions often depend on prior initialization, so the initialization step included in the sequence specification prevents many false alarms in module-level analysis.


\section{Automatically Inferring Contracts}
\label{sec:contractinference}

Large code bases and legacy code often lack contract annotations, and adding them manually at scale is infeasible.
To make our framework applicable to both large code bases and legacy code, automatic inference of contracts during module-level analysis is an essential part of it. 
Two ways to infer contracts are supported: 
First, we use interface specifications as contracts.
Second, we derive additional contracts by leveraging results of previous \Astree{} runs.
Both techniques correspond to the upper part of Fig.~\ref{fig:overview}.
Although inferring contracts from abstract interpretation is not a new idea (see related work in Sec.~\ref{sec:related}), we did not find prior work doing so in module-level analysis.

\subsection{Contracts from Interface Specifications}
\label{sec:interfacespecs}

There are several standards for software interfaces between automotive electronic control units.
Examples are AUTOSAR~\cite{autosar} or ASAM MDX~\cite{asam-mdx}.
These interface descriptions ensure that modules in the automotive industry can be reused in different car models from different carmakers.

When value ranges for the incoming and outgoing values of a given module are defined in the interface descriptions in AUTOSAR, they are automatically parsed into preconditions, postconditions, and array specifications. 
For arrays, this is possible because they are explicitly represented as types with corresponding lengths.
ASAM MDX focuses on messages being passed between modules.
Here, value ranges can be automatically parsed into invariants.

\begin{example}
An AUTOSAR interface can look as follows:
{\small
\begin{verbatim}
  <DATA-CONSTR>
    <SHORT-NAME NAME-PATTERN="{anyName}">
      RangeX
    </SHORT-NAME>
    [...]
    <PHYS-CONSTRS>
      <LOWER-LIMIT INTERVAL-TYPE="CLOSED">
        0.0
      </LOWER-LIMIT>
      <UPPER-LIMIT INTERVAL-TYPE="CLOSED">
        32000.0
      </UPPER-LIMIT>
    </PHYS-CONSTRS>
    [...]  
  </DATA-CONSTR>
\end{verbatim}
}
\noindent	
The above AUTOSAR interface defines a closed interval $[0.0, 32000.0]$ that can be referred to as \texttt{RangeX} in the definition of a variable of a module and yields a function contract like:
{\color{blue}\texttt{[[ requires: x >= 0.0 \&\& x <= 32000.0 ]]}}
\end{example}

Similar specifications exist for curves and maps that are used to adapt the software to the specifics of a given car model.
Leveraging all this information turns out to be a great resource for contracts which drastically reduce the number of false alarms in module-level analysis.

\subsection{Contracts from Abstract Interpretation}
\label{sec:contractsfromastree}

The second way our framework infers contracts is to automatically utilize results from previous runs of \Astree{}.
This relies on the precision of \Astree{}.

\begin{example}
Consider the following function:
\begin{lstlisting}
float foo(float x) {
  return 1.0f / bar(x);
}
\end{lstlisting}
Knowing nothing about the function \texttt{bar()}, \Astree{} reports a potential division by zero in this function as \texttt{bar()} might return zero. However, we might have analyzed the implementation of \texttt{bar()} as part of some other module:
\begin{lstlisting}
float bar(float x) {
  return max(x, 1.0f);
}
\end{lstlisting}
For that function, \Astree{} detects that its return value is always greater than or equal to one yielding the inferred contract {\color{blue}\texttt{[[ ensures: return >= 1.0 ]]}}. Storing this knowledge and reusing it for the analysis of function \texttt{foo()} then helps to prevent the false division by zero alarm in \texttt{foo()}.
\end{example}

Generally, after letting \Astree{} run on a module, we read out abstract values for all output and return messages from functions in the module and store them in a database. 
This corresponds to step ``Process Report'' in  Fig.~\ref{fig:overview}.
The same is done for function parameters of stubbed functions to find out how these functions are used.
Our framework extracts the resulting abstract values via additional directives in our generated verification harness (line~\ref{line:additionalDirectives} in Algorithm~\ref{alg:stubgen}).
In case of multiple outputs for the same variable or function, e.g., due to multiple writers or dynamic binding, these abstract values are combined via least upper bound over all possible results. They can then be used as preconditions in the next iteration and lead to more precise analysis of the functions that use these values.

More formally, let $G$ be the set of global variables, $L$ the abstract domain (a lattice), $M$ the set of modules, $F_m$ the set of functions in module $m \in M$, and $R_m : F_m \times G \rightarrow L$ the function that returns which value a global variable has after running a function from $F_m$. The function $R_m$ returns $\bot$ when the function from $F_m$ does not change the variable. Then, the possible values of a variable $v \in G$ after analysis of module $m$ is calculated as follows:
\begin{equation}
{\rm Value}_m(v) = \bigsqcup_{f \in F_m} R_m(f,v)
\label{eq:valuem}
\end{equation}
$\bigsqcup$ is the least upper bound function. To aggregate results from multiple modules, the same least upper bound function is used:
\begin{equation}
{\rm Value}(v) = \bigsqcup_{m \in M} {\rm Value}_m(v)
\label{eq:value}
\end{equation}
Analogous calculations are done for function parameters Param$(f,p)$ and return values Return$(f)$.

To create contracts based on these abstract values, knowledge about the abstraction is required. 
For an interval domain, the abstraction contains the lower and upper limit of possible values. 
Those can easily be translated to a contract: {\color{blue}\texttt{[[ invariant: x >= lower \&\& x <= upper ]]}} 
For a set domain, the contract is a disjunction of possible values: {\color{blue}\texttt{[[ invariant: x == 0 || x == 10 ]]}}
In the following, the sets of resulting contracts are denoted by $\FC'$ and $\VC'$ for function and variable contracts, respectively.

\begin{algorithm}[t]
\SetKwInOut{Input}{Input}\SetKwInOut{Output}{Output}
\Input{Modules $M$, function contracts $\FC$, variable contracts $\VC$}
\Output{Potential runtime errors in $M$, inferred function and variable contracts $\FC'$, $\VC'$}
\BlankLine
$\FC'$=empty, $\VC'$=empty\;
\label{line:InitInFixpointAlg}
\Repeat{{\rm no more changes in $\FC'$ and $\VC'$}}{
	\ForAll{$m \in M$}{
	  ({\it alarms}, {\it values}) = run \textbf{Algorithm\,\ref{alg:stubgen}}(m, $\FC \oplus \FC'$, $\VC \oplus \VC'$)\;
	  \label{line:CompVerificationInFixpointAlg}
		update Value$_m$, Param$_m$, and Return$_m$ based on {\it values}\;
		\label{line:UpdateModuleValuesInFixpointAlg}
		update Value, Param, and Return\;
		\label{line:UpdateGlobalValuesInFixpointAlg}
		update $\FC'$ and $\VC'$ based on that\;
		\label{line:UpdateContractsInFixpointAlg}
	}
}
\Return ({\it alarms}, $\FC'$, $\VC'$)
\BlankLine
\caption{Runtime Error Analysis with Contract Inference.}
\label{alg:fixpoint}
\end{algorithm}

Algorithm~\ref{alg:fixpoint} shows the overall algorithm for performing module-level runtime error analysis with contract inference.
As input, we start with function contracts~$\FC$ and variable contracts~$\VC$ which both can be manually written or derived from standardized interfaces.
Inferred function contracts~$\FC'$ and inferred variable contracts~$\VC'$ are empty initially (line~\ref{line:InitInFixpointAlg}).
Variables not concerned by the contracts are set to the full range of abstract values.
On each module, Algorithm~\ref{alg:stubgen} is run using the current set of contracts (line~\ref{line:CompVerificationInFixpointAlg}). The operator $\oplus$ denotes the merging of contracts, where only elements not already present on the left-hand side are added.
In line~\ref{line:UpdateModuleValuesInFixpointAlg}, the results of abstract interpretation are used to update Value$_m$, Param$_m$, and Return$_m$ according to Eq.~\ref{eq:valuem}.
In line~\ref{line:UpdateGlobalValuesInFixpointAlg}, the aggregation functions Value, Param, and Return are calculated based on all modules' results according to Eq.~\ref{eq:value}.
Finally, the contracts $\FC'$ and $\VC'$ are derived from the aggregation functions using knowledge about the underlying abstract domains (line~\ref{line:UpdateContractsInFixpointAlg}).
The updated contracts are used to analyze the next module. This process is repeated, as new contracts may be identified that lead to changed results in a subsequent analysis run.
Eventually, this process reaches a fixpoint, as contracts can only be refined by this process (additional and increasingly refined restrictions added via $\oplus$). 
The number of needed iterations depends on the number of modules with effects on each other and on the analysis order.

A database is used to store results for individual modules to enable analysis in a distributed setting.
Every analysis recalculates contracts based on module-level results for all modules (similar to lines~\ref{line:UpdateGlobalValuesInFixpointAlg} and~\ref{line:UpdateContractsInFixpointAlg}) that are retrieved from the database to take changes on the module level into account. 

Note that during development of a module, it is not required to run this entire process after every change. It is sufficient to run the analysis only on the changed module, which leads to an update of the database. This way, the derived contracts may not always be up-to-date, but it is sufficient for addressing immediate issues at this stage. The full check up to the fixpoint will still be done regularly (e.g., in a nightly build), and the database is updated accordingly.



\section{Case Study}
\label{sec:eval}

We implemented the described framework in Scala~\cite{scala} and applied it to three automotive embedded subsystems under development\footnote{The tool is a company-internal development, which we cannot share. The subject systems consist of productive industrial code, which we cannot share, either.}. Table~\ref{tab:systems} shows an overview of their characteristics. The first one is a driver assistant subsystem~(A), the second a braking subsystem~(B), and the third one a cruise control subsystem~(C). System~A is written in~C, systems~B and~C in~\CPP{}. 
System~B contains the highest number of functions and ifs as well as the lowest number of loops.

\begin{table}[t]
\centering
\begin{tabular}{lccc}
\hline
System      & ~Driver Assistance (A)~ & ~Braking (B)~ & ~Cruise Control (C)~ \\
\hline
Language    & C  & \CPP{} & \CPP{} \\
KLOC        & 17 &   6 &   7 \\
\#functions & 66 & 124 &  95 \\
\#ifs       & 70 & 102 &  36 \\
\#loops     & 49 &   1 &  28 \\
\hline
\end{tabular}
\vspace{.5em}
\caption{Characteristics of subject systems.}
\label{tab:systems}
\vspace{-0.35cm}
\end{table}

We apply integration analysis, which analyzes the complete software project where contracts and stubs are not necessary as all implementations are available.
This baseline is called stage~0. 
We then perform module-level analysis as described in Sec.~\ref{sec:approach} with incremental stages of contracts. 
In stage~1, we use module-level analysis only with implicit contracts based on type information such as enum ranges and floats excluding invalid values like \texttt{NaN}. 
In stage~2, we also use automatically inferred contracts as described in Sec.~\ref{sec:contractinference}:
Only for system~B, interface specifications are available and used, but for all three systems, contracts resulting from abstract interpretation up to a fixpoint are used.
In stage~3, we additionally use contracts that were manually written by experts. 
Overall, 6, 0, and 13 contracts were written manually for systems A, B, and C, respectively, whereas 28, 147, and 127 contracts were inferred automatically for those systems.
We use the same set of \Astree{} settings in all four stages.

\begin{table}[t]
\centering
\begin{tabular}{cc|ccccccccc|cc}
\hline
System/~& Total~    & IPA~& ISA~& IRO~& DMZ~& UIV~& UFC~& DCF~& CPP~& ASR  & Definite & ~Coverage~ \\
Stage~  & Alarms~   &     &     &     &     &     &     &     &     &      & Alarms   & [\%]       \\
\hline                                                                                    
A0      &    131    &  21 &  11 &  17 &     &     &  51 &  33 &     &      & 8        & 83.7       \\
\hdashline
A1      &    131    &  24 &  11 &  20 &     &     &  37 &  39 &     &      & 8        & {\bf 90.4} \\
A2      &    108    &  20 &   1 &  11 &     &     &  35 &  41 &     &      & 7        & 89.4       \\
A3      & {\bf 102} &  14 &   1 &  11 &     &     &  35 &  41 &     &      & 6        & 89.4       \\
\hline                                                                                             
B0      &     21    &     &     &  21 &     &     &     &     &     &      & 0        & 64.8       \\
\hdashline
B1      &     92    &   3 &   3 &  61 &   7 &     &     &     &     &  18  & 2        & 93.8       \\
B2      & {\bf 12}  &   1 &     &   8 &   1 &     &     &     &     &   2  & 2        & {\bf 99.3} \\
\hline                                                                                             
C0      &    190    &  22 &   4 & 129 &   6 &  10 &  17 &     &     &   2  & 2        & 75.8       \\
\hdashline
C1      &    304    &  24 &     & 167 &   6 &  61 &   5 &   8 &  19 &  14  & 20       & {\bf 99.8} \\
C2      &    301    &  22 &     & 166 &   6 &  61 &   5 &   8 &  19 &  14  & 18       & {\bf 99.8} \\
C3      & {\bf 173} &   3 &     & 139 &   6 &  12 &   5 &     &     &   8  & 3        & {\bf 99.8} \\
\hline
\end{tabular}
\vspace{.5em}
\caption{Case study results. Alarm classes: IPA = invalid usage of pointers or arrays; ISA = invalid shift argument; IRO = invalid ranges and overflows;
DMZ = division or modulo by zero; UIV = uninitialized variables; UFC = unknown function called; DCF = data and control flow alarms; CPP = C\texttt{++} specific alarms; ASR = failed asserts.}
\label{tab:eval-results}
\vspace{-0.35cm}
\end{table}

Table~\ref{tab:eval-results} shows the results. 
The first column indicates system and stage, and the second column shows the total number of alarms.
The columns in the middle denote the individual alarm classes as a decomposition of the total number of alarms. 
These alarm classes originate directly from the report of \Astree{} and are resolved in the caption of the table.
The last two columns are discussed later.

In all three systems, the total number of alarms increases or stays the same when using module-level analysis (stage~1) in comparison to integration analysis (stage~0). It decreases below the values from stage~0 when additionally using contracts.
The initial increase in alarms is caused by an increase in coverage as the share of reached code.
In systems~A and~B, automatic contract inference in stage~2 suffices to largely eliminate additional false alarms. 
In system~A, the alarms from classes~UFC (unknown-function-called) and~DCF (data-and-control-flow-alarms) originate from the use of compiler specific functions that \Astree{} does not know and from intentional infinite loops that are used while waiting for a reset to handle internal errors.
Without these, the total number of alarms is reduced from 49 for A0 to 26 for A3.
In system~B, the results of stage~2 were so good that no manual contracts had to be added, therefore B3 is not required. 
In system~C, manually written contracts are required for satisfactory results. 
This is due to complex dependencies between modules in this system.

B2 shows a significant reduction in the number of alarms in the class IRO (invalid-ranges-and-overflows) demonstrating how well automatically inferring contracts works when contracts can express the respective property. 
In B1 and B2, we get some alarms from classes that do not occur in B0.
These alarms stem from limitations of the expressiveness of our contracts that are necessary to keep them checkable by \Astree{}.
An example for such a limitation is that we cannot express the relationship between subsequent elements in an array.
When an element is subtracted from the previous element in the array and the result is used as a divisor, then a division by zero occurs if both elements are the same.
Although we observe that subsequent elements are always different, we cannot express this property in a contract and thus cannot avoid the false alarms. 

Alarms from class ASR (failed-asserts) often represent violations of contracts as preconditions and postconditions are encoded by assertions.
This is an additional benefit of our framework as it facilitates easier evaluation of alarms to be true or false alarms.
If a precondition is not fulfilled, the corresponding assertion alarm is close to the calculation of the responsible value.
Without the assertion, another alarm would happen deeper in the function, which is harder to evaluate.

Table~\ref{tab:eval-results} also shows the number of definite alarms.
A \emph{definite alarm} occurs when all abstract values of a variable lead to an alarm. An example for a definite alarm is dereferencing a pointer that is always null at that point as in this example: \texttt{int *p = NULL; int x = *p;}.
This is in contrast to a \emph{possible alarm} where at least one but not all abstract values of a variable lead to an alarm. 
Here, \Astree{} continues the analysis with the remaining values, excluding the values causing the error.
In case of a definite alarm, \Astree{} cannot continue with any remaining value and thus aborts the execution path.
Therefore, developers should fix definite alarms that are followed by other code to increase code coverage.
Note that the number of definite alarms can decrease from stage~0 to stage~3 because the increasing number of contracts can restrict the value range of input variables, which can remove possible and definite alarms. 

Coverage almost always increases for all considered systems. In most cases, some functions are not used in the integrated system, but module-level analysis checks all of them. In other cases, as analysis aborts analysis paths on definite alarms, more subsequent code may not be reached in integration analysis compared to module-level analysis, where each method is called individually from the main loop.
We also manually compared alarms from integration analysis with alarms found in stages~2 and~3 and observed that no relevant alarms were missing, but additional alarms were found in code that was not reached before.

Table~\ref{tab:runtimes} shows the runtimes of integration versus module-level analysis on an Intel Xeon 4~GHz with 4~cores and 32~GB RAM as client and an Intel Xeon 3.2~GHz with 32~cores and 512~GB RAM running the \Astree{} server.
There are big differences between C and \CPP{} systems. The C~system (A) takes 710~seconds for integration analysis, but only 4~seconds (median) per module for module-level analysis. 
For \CPP{}, the relatively high overhead of \CPP{} preprocessing by \Astree{} leads to module-level analysis times of 20~seconds (B) and 48~seconds (C) (median). In integration analysis, this overhead is only due once and the total runtime is only 41~seconds (B) and 157~seconds (C), respectively.
This means the savings when checking an individual module are rather small in the \CPP{} cases, but the advantage of being able to perform analysis early during development when integration is not yet possible still remains -- along with increased coverage. 


\begin{table}[t]
\centering
\begin{tabular}{cp{0cm}|p{0cm}ccccccp{0cm}|p{0cm}c}
\hline
Sys- & & & \multicolumn{6}{c}{Module-level analysis} & & & Integration analysis \\
tem	 & & & median(T)   & avg(T)         & max(T)         & \#outliers(T) & ~$|T|$~ & sum(T) & & & total \\
\hline
A & & & \hphantom{4}4s & \hphantom{1}13s & \hphantom{7}89s & 2 & 18 & \hphantom{1}241s & & &            710s \\
B & & &            20s & \hphantom{1}21s & \hphantom{7}32s & 3 & 45 & \hphantom{7}950s & & & \hphantom{1}41s \\
C & & &            48s &            116s &            789s & 2 & 13 &            1510s & & &            157s \\
\hline
\end{tabular}
\vspace{.5em}
\caption{Runtime comparison. Runtimes for all modules of respective system are in~T.}
\label{tab:runtimes}
\vspace{-0.35cm}
\end{table}

Analysis times of individual modules vary a lot. For example, in system~C, one of the modules takes 789~seconds, which is more than half of the total module-level analysis runtime for this system. This module contains nested loops, which are expensive to analyze when loop unrolling is active in \Astree{}. The analysis time for this module is several times higher than the total integration analysis time, which indicates that a more concrete scenario is checked in the integration case, and thus loops have to be analyzed less often. However, the majority of modules is analyzed within at most two minutes. The number of outliers (according to Tukey's fences~\cite{tukey:book} with k=1.5) is low as shown in the table and gives an indication about the share of modules requiring more processing time.
Conducting this case study also showed that contract inference causes only a low runtime overhead (e.g., 10\% in case of system~B).

In summary, the evaluation of the case study shows that our framework can help developers in their daily work to detect runtime errors early during development.
This is due to low analysis runtime for individual modules, almost no manual effort for writing contracts, and high precision.


\section{Practical Experiences with Large Systems}
\label{sec:largescale}

We describe our experiences from letting developers of different projects use our framework on two large systems: a control software and a software library. 

\subsection{Embedded Control Software}

For a large embedded control software system of approx.\ two MLOC in about 3,000 C~files, the integration analysis of the system ran for more than three days with low precision settings. 
A run with high precision settings for \Astree{} using loop unrollings, partitionings, and relational domains was aborted after several weeks.
By contrast, using module-level analysis via our framework reduces the runtime to 12~hours. 
This could be reduced even further by massive parallelization, which is only possible due to module-level analysis.

Additional to the runtime improvements, our framework made it possible to run \Astree{} with high precision settings. 
The disadvantage of losing context information could be compensated by inferring contracts from abstract interpretation as described in Sec.~\ref{sec:contractsfromastree}: The number of alarms in module-level analysis was reduced by 28\% when using contracts from abstract interpretation. 
The reduced number is still higher than in integration analysis mainly because all code from all modules is covered -- and not only the code that is actually used.
The main benefit of our framework in this case is that the analysis of modules can be executed at an early stage when the system is not yet integrated. 

\subsection{Embedded Software Library}

Another team applied our framework to a foundational library used in multiple automotive software systems.
The software library comprises approx.\ one~MLOC in 2,000 C~files and is organized in modules. 
Each module consists of 10 to 50 compilation units that may share common data, but its external interface only consists of C~functions. 
Our framework analyzes each module individually.
The analysis reaches a high code coverage of more than 90\% in most modules, which is important for a library.
As each module consists of only a few thousand lines of code, \Astree{} can be utilized with high precision settings, e.g., most loops can be fully unrolled. This leads to 10 to 100 reported alarms per module.
The alarms have a short call-stack of at most three calls and hence are easy to review by a human. This is an advantage compared to integration analysis where call stacks are often large and hard to follow.
The share of alarms that should lead to code changes is approx.\ 10\%, which is a remarkably high rate for abstract interpretation in industrial use according to our experience.

Notice that the 10\% alarms that should lead to code changes include so-called \emph{justified} alarms.
An alarm is justified when it may occur by use of a function respecting its contracts.
When considering library verification, a justified alarm is different from a bug in production code.
As an example, consider the following library function:
\begin{lstlisting}
int Curve[10];
int getEntry(int i) {
	return Curve[i];
}
\end{lstlisting}
In the code using the library, this function might always be called correctly, following implicit assumptions. Thus, if the integrated system is analyzed, then the statement \texttt{Curve[i]} never yields an array-out-of-bounds alarm.
However, from a library point-of-view, the above implementation of the function is buggy if used incorrectly. Our framework discovers such issues:
Our driver calls the function with arbitrary input for \texttt{i} by setting the contents of the parameter variable to full range when no contracts exist.

Here, contracts are useful in the interaction of verification engineers and core developers when these roles are distributed over different teams, time zones, or countries: 
Verification engineers can directly write contracts for the implicit assumptions of the library.
This benefits both verification engineers and core developers:
Verification engineers do not encounter alarms with the same root cause again and again.
Also, contracts and the resulting \Astree{} directives are both more convenient and safer than commenting alarms to dismiss them.
Core developers benefit from implicit assumptions being made explicit by the contracts in case they do not change their code to check them.
Both verification engineers and core developers benefit from the automatic detection of future violations of the implicit assumptions being made explicit via contracts.




\section{Lessons Learned}
\label{sec:lessons}

One issue preventing successful use of abstract interpretation in the automotive industry is that developers perceive it as a burdensome activity due to many false alarms.
Oftentimes, alarms are only manually annotated if identified as false alarms, and this process is repeated when parts of the software are reused.
A contributing factor to this issue is that abstract interpretation is only applied to the integrated system in late stages of development.

We identify module-level abstract interpretation with contracts as solution to these problems, because developers get feedback early during development and can fix issues while they still know the code.
To deploy module-level analysis, we learn that central availability and automation is a prerequisite.
Also, contracts are needed to resolve missing contexts of modules.
We identify the contract types of preconditions, postconditions, array specification, sequence specifications, and invariants as a good middle ground in terms of difficulty of automatic inference, provided expressiveness, and needed overhead for analysis in the automotive domain.
These types of contracts are also intuitive for developers such that they can write them manually on newly produced code.
With the described contract inference, they can be automatically obtained for large legacy code bases.

Developers get first results with our framework after a few days because we designed it to be adaptable to new projects with custom extensions to account for the specific architecture models, processes, and conventions used in the projects.
The first results are often already helpful for developers and convince them of the benefits of using abstract interpretation.
Developers appreciate the short runtimes of module-level analysis as they do not block their usual workflows.

When developing and deploying our framework, we also found possibilities for improvements when applying \Astree{} to automotive software. 
Analyzing \CPP{} has a higher overhead compared to C, but every speedup in processing of \CPP{} would be appreciated. 
Also, templates are difficult to statically analyze.
For container classes, it is often unclear how to instantiate them with representative instances. 

Our framework focuses on \Astree{} as the underlying engine for abstract interpretation, although the general ideas behind our framework can also be used with abstract interpretation tools like Frama-C EVA plugin~\cite{DBLP:phd/hal/Buhler17,buhler2017eva}, Polyspace Code Prover~\cite{polyspaceCodeProver}, and TrustInSoft Analyzer~\cite{trustInSoftAnalyzer}.
\Astree{} is highly customizable, and we benefit from a large knowledge base on settings for different projects comprising years of experience in using the tool.


\section{Related Work}
\label{sec:related}

\paragraph*{Contract specification}

The concept of contracts was introduced by Meyer for Eiffel and used for dynamic checks~\cite{meyer:book88}.
SPARK extended Ada with contracts~\cite{carre:triada90} that can be checked statically or at runtime.
Leavens adapted contracts for Larch/\CPP{}~\cite{leavens:oobs96,leavens:fm99} and then co-authored JML, a behavioral interface specification language for Java~\cite{leavens:bsbs99,chalin:fmco06}, which later evolved into OpenJML~\cite{cok:nasafm11}. The latter is still under active development.
Microsoft Research developed Spec\# as a C\# variant with contracts~\cite{leino:laser10,barnett:cacm10}.
They then changed to an API based approach called CodeContracts~\cite{faehndrich:fvoos10,faehndrich:sac10,logozzo:vmcai11}.
Frama-C supports a language called ACSL for specifying contracts, which is embedded into comments and supports predicate logic~\cite{cuoq:sefm12}.
Hatcliff et al.\ give an overview of behavioral specification languages and their commonalities and differences~\cite{hatcliff:comsur12}.
Our contract specification syntax uses similar concepts as all of these publications -- there seems to be a broad consensus of how contracts should look like.




\paragraph*{Contract verification}

The first static contract checking tool for JML was called ESC/Java~\cite{flanagan:pldi02}. It translated source code and specifications into verification conditions and passed them to a theorem prover. Its successor \mbox{ESC/Java2} offered support for more JML specifications~\cite{cok:cassis05,chalin:fmco06} and Java 1.4. OpenJML is the current JML verification tool~\cite{cok:nasafm11}. It is based on OpenJDK, which enables easy adaptation to new language versions.
The static checker for Spec\# is Boogie, which has an own intermediate language making it usable for multiple languages~\cite{barnett:fmco06}.
A checker for CodeContracts exists in the form of cccheck/\allowbreak Clousot~\cite{faehndrich:fvoos10,logozzo:vmcai11}.
Similar to us, they transform pre- and postconditions into assume/assert calls and then apply their abstract interpretation engine. Frama-C is capable of checking certain contracts written in ACSL~\cite{cuoq:sefm12,kirchner:frama-c}. It is also based on abstract interpretation~\cite{DBLP:phd/hal/Buhler17,buhler2017eva}.
Despite the availability of these approaches and tools, formal methods are rarely used when developing automotive software~\cite{DBLP:conf/issta/AltingerWS14}.



\paragraph*{Contract inference}

Abstract interpretation~\cite{cousot:popl77} can also be seen as a technique to infer invariants, as it iterates up to a fixpoint. 
We use this technique to find postconditions.
Daikon uses dynamic analysis and expression patterns to infer likely invariants from Java code~\cite{ernst:icse99}.
Ammons et al.\ use dynamic analysis to infer API usage protocols from code~\cite{ammons:popl02}.
Arnout et al.\ investigate .NET code for implicit contracts, which they find in form of exceptions and documentation~\cite{arnout:fmco02}. 
F\"{a}hndrich and Logozzo infer preconditions in simple cases -- when the respective variables are not modified~\cite{faehndrich:fvoos10}.
Wei et al.\ use dynamic analysis to infer contracts from Eiffel code~\cite{wei:icse11}.
Cousot et al.\ describe how to use abstract interpretation to infer necessary preconditions~\cite{cousot:vmcai11,cousot:vmcai13} (implemented in cccheck/Clousot~\cite{carr:tse17}). 
Our framework can use contracts obtained by any of these approaches.

\paragraph*{Modular verification}

Alur et al.\ present a theoretical approach for automatically constructing abstractions of modules for modular verification of finite-state systems~\cite{alur:concur99}.
Chaki et al.\ propose an approach for modular verification of C~code against finite state machines using procedure abstractions~\cite{chaki:tse04}.
Cohen et al.\ describe a verifying C compiler that uses Boogie to prove correctness in a modular way~\cite{cohen:tphols09}. 
Letan et al.\ show how the properties of a composition of proven components can be checked based on abstract state specifications~\cite{letan:fm18}.
By contrast, our framework works without manually writing formal specifications.




\section{Conclusion}
\label{sec:conclusion}

We presented a framework to detect runtime errors in automotive software with low manual effort. 
Our framework enables analysis early during development which has the following benefits: 
First, developers get feedback quickly and can fix problems when they still deeply understand the software. 
Second, the analysis runs faster on an individual module in comparison to the entire software system. 

Our framework combines abstract interpretation with contracts.
Contracts provide detailed information about the behavior of other modules that might not yet be implemented.
They can be written manually or derived fully automatically.
We use the abstract interpretation tool \Astree{} under the hood and presented how contracts can be utilized by \Astree{} to improve precision.

We evaluated our framework on several automotive software systems of different size.
Our case study showed that the number of found alarms, the runtime, and the code coverage can be greatly improved.
For the domain of braking, our framework decreased the number of alarms by approx.\ 50\% while increasing code coverage by approx.\ 50\%.
Our framework is used by developers in their daily work on projects with over a million lines of code. 
The feedback so far has been very positive, and the rollout to other projects is planned. 


\bibliographystyle{splncs04}
\bibliography{paper}

\end{document}